# Study on the unfolding algorithm for D-T neutron energy spectra measurement using recoil proton method[*]

WANG Jie(王洁)[1]   LU Xiao-Long(卢小龙)[1,2]   YAN Yan(严岩)[1]   WEI Zheng(韦峥)[1]

WANG Jun-Run(王俊润)[1]   RAN Jian-Ling(冉建玲)[1]   HUANG Zhi-Wu(黄智武)[1]

LAN Chang-Lin(兰长林)[1,2]   YAO Ze-en(姚泽恩)[1,2;1)]

[1] School of Nuclear Sciences and Technology, Lanzhou University, Lanzhou 730000, China

[2] Engineering Research Center for Neutron Application, Ministry of Education, Lanzhou University, Lanzhou 730000, China

**Abstract**: A proton recoil method for measuring D-T neutron energy spectra using polyethylene film and Si (Au) surface barrier detector was presented. An iteration algorithm for unfolding the recoil proton energy spectrum to the neutron energy spectrum was investigated. The response matrices $R$ of polyethylene film at 0 degree and 45 degree were obtained by simulating the recoil proton energy spectra from the mono-energetic neutron using MCNP code. Under an assumed D-T neutron spectrum, the recoil proton spectra from polyethylene film at 0 degree and 45 degree were also simulated using MCNP code. Based on the response matrices $R$ and the simulated recoil proton spectra at 0 degree and 45 degree, the unfolded neutron spectra were respectively obtained using the iteration algorithm, and compared with the assumed neutron spectrum. The results show that the iteration algorithm method can be applied to unfold the recoil proton energy spectrum to the neutron energy spectrum for D-T neutron energy spectra measurement using recoil proton method.

**Key words**: D-T neutron source, recoil proton energy spectra, D-T neutron energy spectra, iteration algorithm method, MCNP code

**PACS**: 29.25.Dz, 29.30.Ep, 29.30.Hs

## 1 Introduction

In 1988, a $3.3\times10^{12}$ n/s neutron generator (ZF-300) based on T(d, n)$^4$He (D-T) reaction with a rotating target had been built at Lanzhou University [1]. It had been applied in the research fields of nuclear data measurements, radiation hardening and radioactive breeding [2]. A higher intensity D-T neutron generator is being developed at our laboratory. In the applications of D-T neutron generator, neutron energy spectrum is one of the most important parameters. In previous investigations, a mathematical method had been developed to calculate the energy spectrum from D-T reaction in a thick tritium-titanium target for the incident deuteron beam in energy lower 1.0MeV[3]. In addition, a Monte-Carlo simulation research and an experimental measurement using nuclear emulsion detector also had been carried out for the neutron energy spectrum of ZF-300 D-T neutron generation [4]. However, because of the energy resolution of nuclear emulsion detector is poor, the agreement between simulation results and experimental data is not very good.

In recent years, scintillation detector has been used to measure D-T fast neutron energy spectrum[5-7]. However, the scintillation detector need a complex electronic system to distinguish the neutron signal and γ-ray signal[8]. In order to avoid the use of complex electronic system, we put forward a proton recoil method for measuring D-T neutron energy spectrum using polyethylene film

[*]Supported by the National Natural Science Foundation of China (11375077) and the National Natural Science Foundation of China (21327801).

1) E-mail: zeyao@lzu.edu.cn

and Si(Au) surface barrier detector. The iteration algorithm for unfolding the recoil proton spectrum to obtain the neutron energy spectrum will be investigated in this paper.

## 2  Principle

The basic principle for measuring D-T fast neutron spectrum using recoil proton method is showed in Fig.1. Fast neutrons from D-T reaction at $\theta$ direction are collimated to bombard on polyethylene (PE) film. A thick depletion layer Si (Au) surface barrier detector is installed at $\alpha$ direction to measure the recoil proton energy spectrum. The measured recoil proton energy spectrum will be unfolded to obtained the fast neutron energy spectrum.

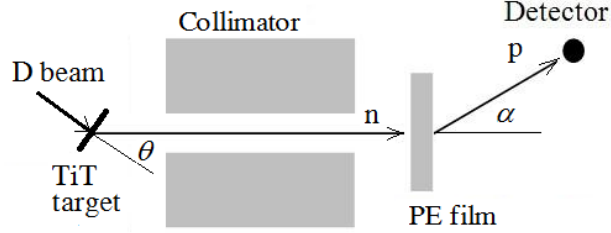

Fig.1 The principle of D-T neutron energy spectra measurement

using recoil proton method

## 3  Unfolding algorithm

The iteration algorithm based on SAND-II method[9,10] will be used to unfold the recoil proton energy spectrum to the fast neutron energy spectrum. Let the matrix $N$ representing the recoil proton energy spectrum data achieved from experiment,

$$N = \begin{bmatrix} N_1 \\ N_2 \\ \vdots \\ N_n \end{bmatrix}, \tag{1}$$

where, $n$ is the energy bin number of the recoil proton spectrum data, $N_n$ is the proton count in $n$-th bin, $N$ should be the linear superposition of all recoil proton energy spectra produced by the monoenergetic neutrons in the incident neutron beam bombarding on PE film. It is assumed that the number of the mono-energetic neutron sources with different neutron energy is $m$. The recoil proton spectrum data produced by every mono-energetic neutron on PE film will conduct a column matrix. All of recoil proton spectrum data will form a matrix $R$ with $n$- lines and $m$- column. The matrix $R$, which is called as response matrix or response function of the polyethylene film, can be expressed by the following equation,

$$R = \begin{bmatrix} R_{11} & R_{12} & \cdots & R_{1m} \\ R_{21} & R_{22} & \cdots & R_{2m} \\ \vdots & \vdots & \ddots & \vdots \\ R_{n1} & R_{n2} & \cdots & R_{nm} \end{bmatrix}, \tag{2}$$

Where, $n$ is the energy bin number of the recoil proton spectrum data, $m$ is the number of the mono-energetic neutron sources. According to the principle of linear superposition, the relationship between the matrix $N$ and the matrix $R$ can be written as the following equation,

$$\begin{bmatrix} N_1 \\ N_2 \\ \vdots \\ N_n \end{bmatrix} = \begin{bmatrix} R_{11} & R_{12} & \cdots & R_{1m} \\ R_{21} & R_{22} & \cdots & R_{2m} \\ \vdots & \vdots & \ddots & \vdots \\ R_{n1} & R_{n2} & \cdots & R_{nm} \end{bmatrix} \cdot \begin{bmatrix} \phi_1 \\ \phi_2 \\ \vdots \\ \phi_m \end{bmatrix}, \quad (3)$$

where, $\phi_1$, $\phi_2$ ... $\phi_m$ are the coefficients of linear superposition. They represent the relative contribution of different energy neutron in the incident neutron beam to recoil proton spectra (N). They can be regarded as the neutron energy spectrum data. However, it is difficult to solve accurately the equation (3) because the $R$ ($n \times m$) is an ill-conditioned matrix.

The iteration algorithm will be used to obtain the neutron energy spectrum based on recoil proton spectrum. Equation (3) can be derived into the following form

$$N_i = \sum_{j=1}^{m}(R_{ij} \cdot \phi_j) = \sum_{j=1}^{m}\{R_{ij} \cdot \exp(\ln\phi_j)\} \quad i=1,2,\cdots,n, \quad (4)$$

where, $R_{ij}$ is the element of the polyethylene film response matrix $R$, $\phi_j$ is neutron count in $j$-th energy bin. Under an assumed initial neutron energy spectrum $\phi_j^{(1)}$, $\ln(N_i)$ in equation (4) can be expanded into the following TAYLOR series truncating which is ignored the high order term after the second term,

$$\ln N_i = \ln N_i^{(1)} + \sum_{j=1}^{m} w_{ij}^{(1)}(\ln\phi_j - \ln\phi_j^{(1)}) + \cdots \quad i=1,2,\cdots,n, \quad (5)$$

where,

$$N_i^{(1)} = \sum_{j=1}^{m}\{R_{ij} \cdot \exp(\ln\phi_j^{(1)})\}, \quad (6)$$

$$w_{ij}^{(1)} = \frac{R_{ij} \cdot \exp(\ln\phi_j^{(1)})}{N_i^{(1)}}. \quad (7)$$

In the iteration algorithm, $\chi^2$ is defined to restrain the iteration results written as

$$\chi^2 = \sum_{i=1}^{n}\{(\ln N_{0i} - \ln N_i) \cdot \frac{1}{\rho_i^2}\}, \quad (8)$$

where, $N_{0i}$ is the measured proton spectrum data in $i$-th energy bin, $\rho_{0i}$ is the relative standard deviations of the $N_{0i}$, $\rho_{0i} = \frac{1}{\sqrt{N_{0i}}}$. To find the minimum valve of $\chi^2$ in equation (4), let

$$\frac{\partial \chi^2}{\partial \ln N_i} = 0, \quad (9)$$

then, an iterative equation for computing the neutron energy spectrum can be derived as [11],

$$\ln \phi_j^{(k+1)} - \ln \phi_j^{(k)} = \frac{\sum_{i=1}^{n}\left(\ln N_{0i} - \ln \sum_{j=1}^{m}[R_{ij} \cdot \exp(\ln \phi_j^{(k)})]\right) \cdot \frac{w_{ij}^{(k)}}{\rho_{0i}^2}}{\sum_{i=1}^{n} \frac{w_{ij}^{(k)}}{\rho_{0i}^2}}, \quad (10)$$

Where, $\phi_j^{(k)}$ and $\phi_j^{(k+1)}$ are respectively the neutron spectrums corresponding to ($k$)-th iteration step and ($k+1$)-th iteration step. After selecting an initial input neutron spectrum $\phi_j^{(1)}$, the iterative step will be repeated until the iterative spectra data in every energy interval satisfy the following criterion

$$\left|\phi_j^{(k+1)} - \phi_j^{(k)}\right| \leq \delta, \quad (11)$$

Where, $\delta$ is set to a small value to control the precision of iterative result. The last iterative data will be regarded as the solution of neutron energy spectrum.

## 4 Simulation and Test

In this section, a simulation method and the simulated recoil proton spectra data are used to verify the feasibility of above iterative method. Firstly, the response matrix $R$ of the mono-energetic neutron at the polyethylene film should be confirmed according to the above iteration algorithm.

### 4.1 Response matrix $R$

The response matrix data should be measured using a series of mono-energetic neutron sources. In fact, the experimental measurements of the response matrix data are unable to be completed because of the lack of the mono-energetic neutron sources. As we all knows, the elastic collision cross section data between neutron and hydrogen nuclei is complete and reliable. In this investigation, the response matrix data will be determined by Monte-Carlo simulation using MCNP code [12].

To simulate the response matrix $R$, a MCNP model is established. As shown in Fig.1, let a mono-energetic neutron barrow beam with a diameter of 2um to incident on the polyethylene film. The thickness of the polyethylene film is selected as 20mg.cm$^{-2}$. According to the energy region of D-T neutron energy spectra, the energy change range of the mono-energetic neutron beam is set to 1MeV to 16MeV with a step of 0.1MeV. F4 card in MCNP code is layouted at α degree direction to record the recoil proton energy spectra. The energy range and the energy interval of the recoil proton energy spectra are selected to 1MeV-17MeV and 0.1MeV, respectively. Type data of the recoil proton energy spectra are shown in Fig.2 (a) and in Fig.2 (b) corresponding to α=0° and α=45° direction, respectively. The response matrixes of PE film with the thickness of 20mg.cm$^{-2}$ corresponding to α=0° and α=45° direction can be respectively made up with the simulated results of recoil proton spectra. The response matrix corresponding to arbitrary angle can also be structured in same simulation method.

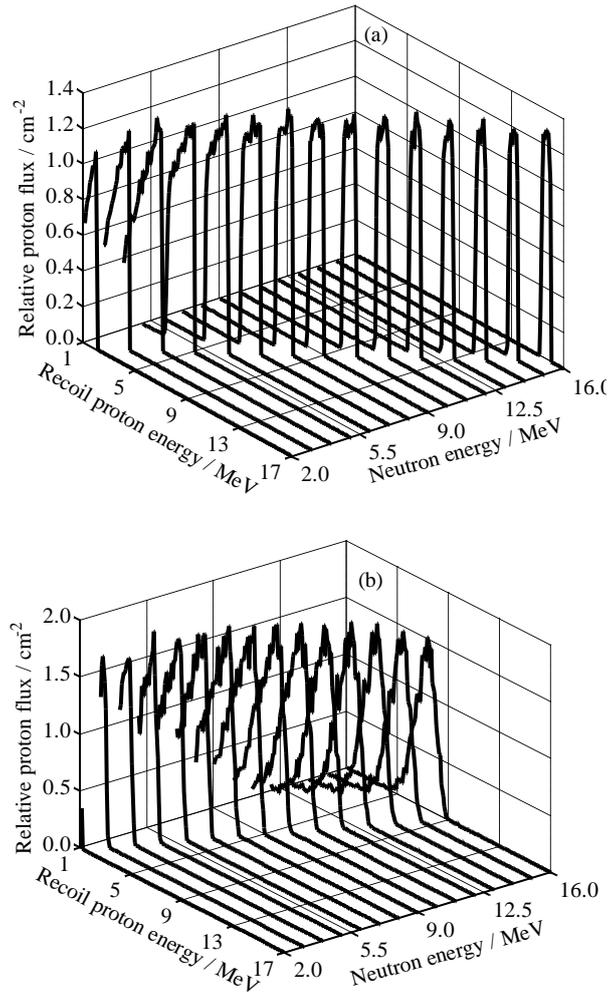

Fig.2 Type data of the recoil proton energy spectra,(a)α=0° ,(b) α=45°

## 4.2 Test method and results

In order to verify the feasibility of above iterative method, a MCNP simulation model is built. As shown in Fig.1, a D-T neutron beam with a diameter of 2um is assumed to incident on PE film with a thickness of 20mg.cm$^{-2}$. Fig.3 shows the energy spectrum of D-T neutron beam, which is based on the previous calculation result [3]. F4 card in MCNP code is layouted at α=0 and α=45 degree direction to record the energy spectra of the recoil proton. The energy interval of the recoil proton energy spectra is also set to 0.1MeV. The simulation results are shown in Fig.4.

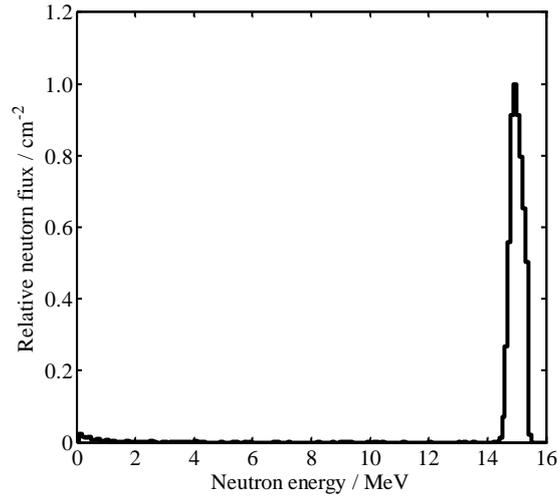

Fig.3 The incident neutron energy spectrum

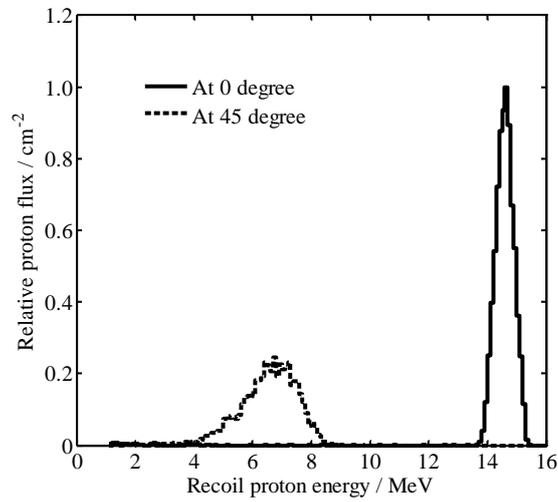

Fig.4 The recoil proton energy spectrum at α=0 and α=45 degree

According to the iterative algorithm shown in equation (10), a computer program is developed using $C^{++}$ code. The response matrix($R$) data in Fig.2 and the recoil proton energy spectrum data in Fig.4 at 0 and 45 degree are respectively input into the program to obtain the neutron energy spectrums by the iterative algorithm. $\delta$ value is set as $10^{-5}$. The unfolded neutron energy spectra using the recoil proton energy spectrum data corresponding to the detector at α=0 (after 12th iteration steps) and α=45 (after 150th iteration steps) degree are respectively shown in Fig.5 and Fig.6. The assumed incident neutron spectrum is also shown in Fig.5 and Fig.6 for comparison.

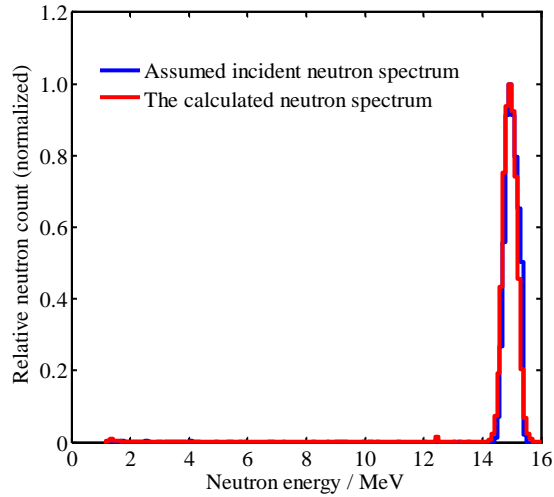

Fig.5 The comparison between the unfolded neutron spectrum corresponding to the detector at α=0 degree and the assumed incident neutron spectrum (color online)

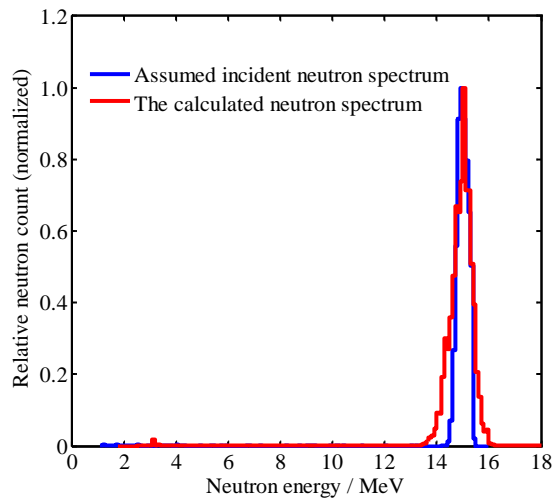

Fig.6 The comparison between the unfolded neutron spectrum corresponding to the detector at α=45 degree and the assumed incident neutron spectrum (color online)

Fig.5 shows that the unfolded neutron spectrum has a perfect match with the assumed incident neutron spectrum while the recoil proton energy spectrum corresponding to the detector at α=0 degree is employed. The full width at half maximum（FWHM）of the incident neutron spectrum and the unfolded neutron spectrum are respectively 0.542MeV and 0.536MeV. From Fig.6, the FWHM of the unfolded neutron spectrum for using the recoil proton energy spectrum corresponding to the detector at α=45 degree is 0.542MeV which is agree with the FWHM of the incident neutron spectrum. However, the width of the unfolded neutron spectrum in Fig.6 is significantly greater than the incident spectrum in 30% peak count below. One reason may be that response matrix($R$) and recoil proton spectrum themselves are broadening, another reason may be that the iteration algorithm still need some optimization method to improve, then above broadening could be accepted in the experimental measurement on D-T fast neutron energy spectrum.

## 5  Conclusions

A proton recoil method for measuring D-T neutron energy spectrum using polyethylene film and Si (Au) surface barrier detector is proposed. The unfolding investigation results show that the developed iteration algorithm can be applied to obtain better neutron energy spectrum. The unfolding neutron energy spectrum for locating detector at α=0 degree is obviously better than the unfolding neutron energy spectrum for locating detector at α=45 degree. However, when the detector is located at α=0 degree, the collimated fast neutron beam will directly incident on Si detector and lead to the damage of the detector. On the other hand, The protons and alpha particles from ( n, p ) and ( n, α ) reactions produced by fast neutron in Si detector will cause an interference for the recoil proton energy spectrum measurement. Based on the above reasons, Si detector should be placed in the shadow zone of collimator to achieve good shielding for Si detector (See Fig.1). It may be a good choice to located the Si detector at α=45 degree.

The main purpose of this research is to confirm the practicability of iteration algorithm, the energy resolution of Si detector is not considered when it is employed to measure the recoil proton energy spectrum. The energy resolution of Si detector with 2000μm depletion layer is lower than 1% for $Am^{241}$ 5.48MeV alphas particle. A better energy resolution should be obtained for the proton with the energy region of 4MeV-16MeV.